\documentclass[11pt]{article}
\usepackage{amsmath}
\usepackage{amssymb}
\usepackage{amsbsy}
\usepackage{ifthen}
\usepackage{color}
\usepackage{graphicx}
\usepackage{float}
\usepackage{enumitem}
\usepackage{natbib}
\usepackage{setspace}
\usepackage{amsthm}
\usepackage{fullpage}
\usepackage[utf8]{inputenc}
\usepackage{tkz-graph}
\usetikzlibrary{arrows,calc}
\usepackage{subfig}
\usepackage{algpseudocode}
\usepackage{algorithm}

\newcommand{\ba}{\begin{eqnarray}}
\newcommand{\ea}{\end{eqnarray}}
\newcommand{\nn}{\nonumber}

\title{International Trade: a Reinforced Urn Network Model}
\author{Stefano Peluso \footnote{Corresponding author. Universit\`a della Svizzera Italiana, Swiss Finance Institute,
  E-mail: stefano.peluso@usi.ch. Via Giuseppe Buffi 13 CH-6904 Lugano. Tel.: +41 (0)58 666 44 95, Fax: +41 (0)58 666 46 47} 
\and Antonietta Mira \footnote{Universit\`a della Svizzera Italiana,  
InterDisciplinary Institute of Data Science and Institute of Finance, 
E-mail: antonietta.mira@usi.ch}
\and Pietro Muliere \footnote{Bocconi University, E-mail: pietro.muliere@unibocconi.it}
\and Alessandro Lomi \footnote{Universit\`a della Svizzera Italiana, InterDisciplinary Institute of Data Science and Institute of Management, Email: alessandro.lomi@usi.ch}}

\begin{document}
\maketitle

\begin{abstract}
We propose a unified modelling framework that theoretically justifies the main empirical regularities characterizing the international trade network. Each country is associated to a Polya urn whose composition controls the propensity of the country to trade with other countries. The urn composition is updated through the walk of the Reinforced Urn Process of \citet{MSW00}. The model implies a local preferential attachment scheme and a power law right tail behaviour of bilateral trade flows. Different assumptions on the urns' reinforcement parameters account for local clustering, path-shortening and sparsity. Likelihood-based estimation approaches are facilitated by feasible likelihood analytical derivation in various network settings. A simulated example and the empirical results on the international trade network are discussed.
\end{abstract}

%%%%%%%%%%%%%%%%%%%%%%%%%%%%%%%%%%%%%%%%%%%%%%%%%%%%%%%%%%%%%%%%%%%%%%%%%%%%
%%%%%%%%%%%%%%%%%%%%%%%%%%%%%%%%%%%%%%%%%%%%%%%%%%%%%%%%%%%%%%%%%%%%%%%%%%%%
%%%%%%%%%%%%%%%%%%%%%%%%%%%%%%%%%%%%%%%%%%%%%%%%%%%%%%%%%%%%%%%%%%%%%%%%%%%%
%%%%%%%%%%%%%%%%%%%%%%%%%%%%%%%%%%%%%%%%%%%%%%%%%%%%%%%%%%%%%%%%%%%%%%%%%%%%
%%%%%%%%%%%%%%%%%%%%%%%%%%%%%%%%%%%%%%%%%%%%%%%%%%%%%%%%%%%%%%%%%%%%%%%%%
%%%%%%%%%%%%%%%%%%%%%%%%%%%%%%%%%%%%%%%%%%%%%%%%%%%%%%%%%%%%%%%%%%%%%%%%%%%%
%%%%%%%%%%%%%%%%%%%%%%%%%%%%%%%%%%%%%%%%%%%%%%%%%%%%%%%%%%%%%%%%%%%%%%%%%%%%
%%%%%%%%%%%%%%%%%%%%%%%%%%%%%%%%%%%%%%%%%%%%%%%%%%%%%%%%%%%%%%%%%%%%%%%%%%%%
\section{Introduction}\label{IntroSection}
\subsection{Related literature and motivation}
Applications of complex network analysis to international trade data have produced a large number of stilyzed facts still lacking unified theoretical justification. Representing international trade relationships as a network with countries as nodes is not novel (\citealt{SK79},\citet{B81}). Research on the statistical and topological properties of the network of international trade relations among countries is more recent. \citet{SB03}, \citet{GL04} and \citet{GL05} analysed the binary version of the international trade network, in which an edge exists between  pairs of vertices representing two countries whenever a positive trade flow between them is observed. The binary representation of trade relations clearly loses crucial information coming from heterogeneous relations that distinguish among major and minor importers and exporters. To overcome this limitation, \citet{BMM07}, \citet{FRS08} and \citet{FRS09} focus on weighted networks of international trade, in which to the edge representing the trade flow is associated a weight proportional to the magnitude of the trade flow existing between the two countries. 

The statistical properties analysed in Bhattacharya et al. (2007) suggest a distribution of node strength that is highly skewed to the right and that may be well approximated by a lognormal distribution. This feature implies that few countries with many trader relations coexist with a large number of countries trading only with a limited number of partner countries. The tendency of more popular vertices in the network to be more likely to attract other vertices is known as \textit{preferential attachment} (\citealt{S55,BA99}). The main reason for scientific interest in preferential attachment stems from its interpretation as a network mechanism capable of generating power law distributions. As a mechanism, preferential attachment is valuable because it generates a long-tailed distribution following a Pareto distribution or power law in its tail. Power law distributions have been shown to characterize a large number of empirical phenomena (\citealt{N06}). Examples include the species distributions, the distribution of the size of cities, of the wealth of extremely wealthy individuals (\citealt{S55}), the number of citations received by scientific publications (\citealt{P76}), and the number of links to pages on the World Wide Web (\citealt{BA99}). We reframe the preferential attachment mechanism of \citet{BA99} in terms of a Polya urn attached to the network, in which: (i) different types of balls correspond to different vertices; (ii) each type of ball may be extracted with probability proportional to the degree (or size, in case of weighted networks) of its associated vertex, and (iii) once a vertex is extracted, it is linked with an edge to a newly introduced vertex.
 
The objective of the present paper is to introduce a flexible modelling framework that is able to capture the main empirical regularities of the weighted network representing the international trade relations. In particular, our network model has the following five constitutive properties: a) In line with the preferential attachment scheme, the popularity of each country (vertex) is positively related to its in-strength, that is the aggregate weight of the edges pointing to the vertex (total imports); b) instead of a single Polya urn attached to the network, as in \citet{BA99}, a system of Polya urns is assumed. Each urn is associated to a country and contains balls corresponding to all or part of the remaining countries; c) a Reinforced Urn Process (RUP, \citealt{MSW00}) moves from country to country (from urn to urn) and through its movement it reinforces the vertices with balls corresponding to the edges traversed, therefore reinforcing the bilateral trade relationships. In this way the preferential attachment mechanism is a local phenomenon, since the popularity of a country may depend on the current position of the RUP, so that a country can be more or less attractive, depending on which country it is trading with; d) The urn construction implies a power law trade flow (edge weight) distribution. In particular, the export flow (weight)
associated to a single directed edge, conditional on all other weights, is a random variable shown to follow a Yule-Simon distribution. Its decay rate in the right tail is inversely proportional to the weights of the other edges starting from the same vertex, creating a competition among edges. This means that the export flow competes with the other trade flows initiated by the same country; e) The traditional preferential attachment model of \citet{BA99} is a generative network model: specification of microdynamic laws facilitates the generation of the networks, but prevents a macroanalysis of the network structure. On the other hand, Exponential random graph models (see \citealt{R07} for a basic discussion) parameterize the network structure, without any hint on the local laws generating the network. We are able to derive the likelihood for directed and undirected, weighted and unweighted networks, so that likelihood-based inference may be performed relatively easily, and still the mechanism driving the motion of the RUP clarifies how to generate the networks.

The basic model is introduced for directed weighted networks, then considerations on equivalence classes of networks allow us to extend the modeling framework to directed unweighted and undirected unweighted networks. We emphasize that, relative to the basic preferential attachment model of \citet{BA99}, the proposed model is notable for its flexibility. As we discuss elsewhere in the paper, different assumptions on the reinforcement parameters governing the RUP process account for various empirical regularities of the international trade network, such as degree distributions and strength distributions skewed to the right, negative assortativity, path-shortening and global sparsity. Also, 
unlike the model of \citet{BA99}), our model may be evaluated or simulated from the likelihood, thus supporting the development of likelihood-based estimation approaches.

%%%%%%%%%%%%%%%%%%%%%%%%%%%%%%%%%%%%%%%%%%%%%%%%%%%%%%%%%%%%%%%%%%%%%%%%%%%%
%%%%%%%%%%%%%%%%%%%%%%%%%%%%%%%%%%%%%%%%%%%%%%%%%%%%%%%%%%%%%%%%%%%%%%%%%%%%
%%%%%%%%%%%%%%%%%%%%%%%%%%%%%%%%%%%%%%%%%%%%%%%%%%%%%%%%%%%%%%%%%%%%%%%%%%%%
%%%%%%%%%%%%%%%%%%%%%%%%%%%%%%%%%%%%%%%%%%%%%%%%%%%%%%%%%%%%%%%%%%%%%%%%%%%%
\subsection{Data description}
International trade provides an ideal context to develop statistical models for networks because economic globalization is frequently interpreted as an ongoing process of greater interdependence among countries and their citizens (\citealt{F03}). The coexistence of positive feedback mechanisms making highly connected countries ever more connected and poorly connected countries ever less connected is an issue of particular concern for the study of international trade and economic development (\citet{R01}). The Direction of Trade (DOT) data set that we use in this paper is available from 1948 to 2000 for 186 countries
The Direction of Trade (DOT) data set is available from 1948 to 2000 for 186 countries. The primary source is the International Monetary Fund DOT Yearbook, in the free version available from the Economics Web Institute. We focus on annual time series of countries' exports and imports, by partner countries, in millions of current-year US dollars. The data are grouped by pairs of countries. Any two countries $x$ and $y$ give rise, in any given year, to two trade flows: exports from $x$ to $y$ and exports from $y$ to $x$. DOT data are based on reports generated by the International Monetary Fund 
member states. However, some of the data for non-reporting and slow-reporting countries are derived based on reports of partner countries.

\begin{figure}[ht]
\begin{center}
\caption{Vertex degree (left) and strength (right) versus the corresponding upper tail probability for 2000 data. The straight lines suggest a right tail behaviour well approximated by the Pareto law.}
\includegraphics[width=15cm]{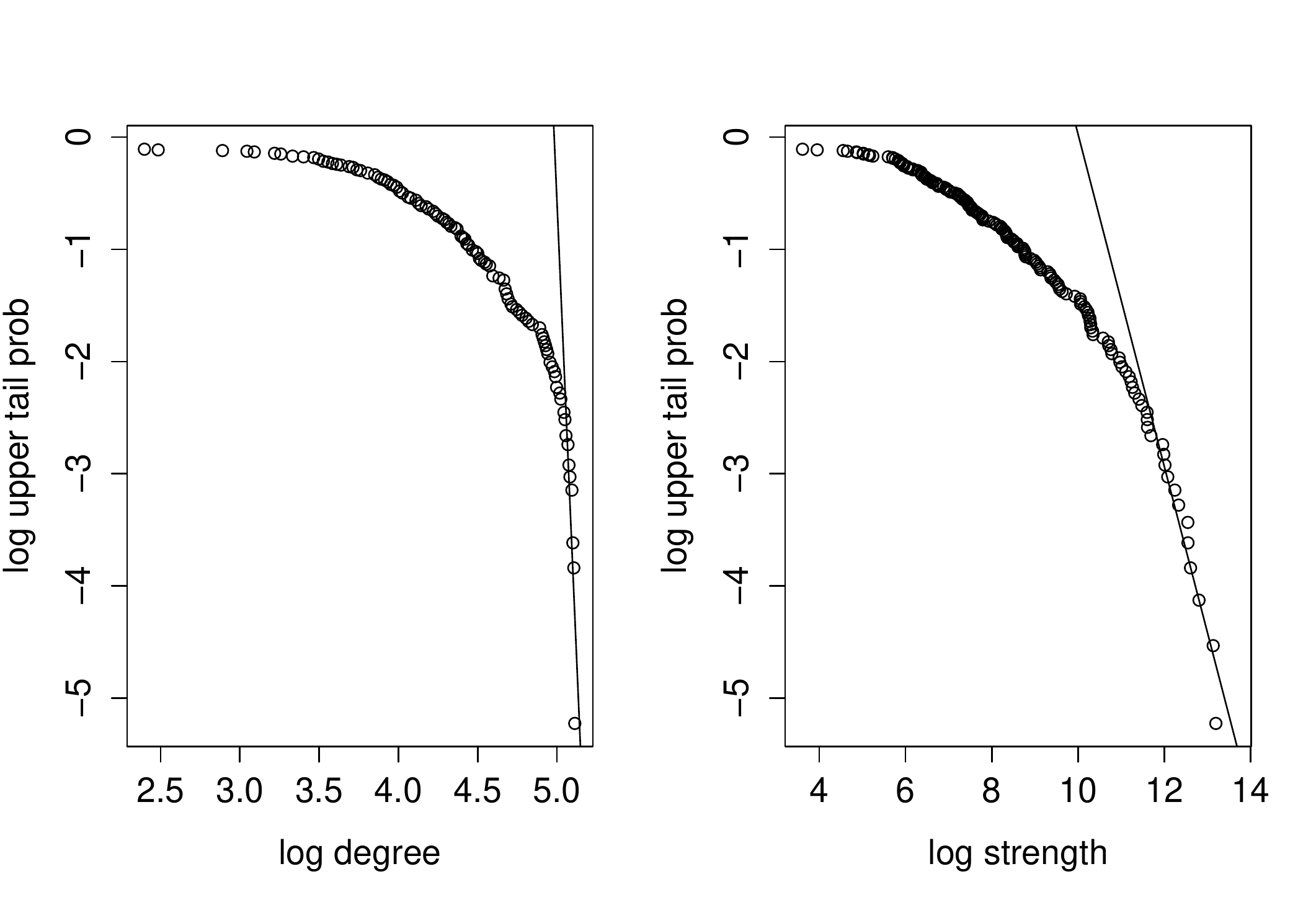}
\label{FigIntro}
\end{center}
\end{figure}

Figure \ref{FigIntro} (left) plots in logarithmic scales for data in 2000, for both the axes, the vertex degree versus the upper tail probability, that is the empirical probability of observing a vertex with higher degree (of observing a country with a higher number of partners). A straight line would indicate a distribution that can be well approximated as a Pareto distribution, but the increasing slope (in absolute terms) suggests that the Pareto approximation can work well only locally. Restricting the analysis to the right tail, the network shows a linear behaviour better approximated by the Pareto law. The simple linear regression line added to the plot is computed on a chosen Pareto scale parameter of 156 (minimum vertex degree) and suggests an Ordinary Least Squares tail index of 32.84. Furthermore, the degree distribution exhibits an empirical positive skewness of 0.2617, suggesting the presence of few countries with a high number of partners, and a larger number of countries with few partners.

Similarly, when we consider the weighted version of the network, taking into account the amounts of bilateral trade flows, we analyse the vertex strength distribution, plotted in Figure \ref{FigIntro} (right): again we note how in log-log scale the whole distribution is far from a Pareto distribution, but, restricting to the right tail, to vertices with strength above 110000 (in millions of current US dollars), a Pareto law with estimated tail index 1.481 can well approximate the tail behaviour. Coherently with \citet{BMM07}, the positive skewness of the strength distribution is a very pronounced phenomenon (in our data estimated to be 4.9443). It highlights the existence of a preferential attachment mechanism producing network centralization, with  a small number of countries accounting for a substantial portion of the total volume of trade. In fact, relative to the full network of international trade, the subnetwork obtained by deleting the vertices corresponding to the 10 largest exporters shows a total traded amount that decreases by 85\%. Furthermore, the sparsity of the subnetwork increases, since the proportion of edges with positive weight (out of the total number of potential edges) decreases from 41.39\% to 35.90\%. 

The average shortest path length, measured as the average number of edges separating any two nodes in the network, is 1.98, clearly exhibiting the so-called \textit{small-world} behaviour (\citealt{WS98}), which implies that the network has an average topological distance between the various nodes increasing very slowly with the number of nodes (logarithmically or even slower), despite showing a large degree of local interconnectedness typical of more ordered lattices. The weighted network has an average weighted local clustering coefficient of 0.9564, computed as in \citet{B04}. This coefficient is a measure of the local cohesiveness that takes into account the importance of the clustered structure on the basis of the amount of trade flows. See \citet{OP09} for an alternative generalization of the clustering coefficient to weighted networks. The comparison of the weighted clustering coefficient with a much lower average topological local clustering coefficient (which does not account for different edge weights) of 0.2047 reveals a network in which the interconnected triplets are more likely formed by the edges with larger weights. In this case clusters have a major effect in the organization of the network because the largest part of the trade flows is occurring on edges belonging to interconnected triplets. Finally, coherently with the empirical evidence in \citet{SK79}, \citet{B81} and \citet{FRS08}, both in the unweighted and, to a lesser extent, in weighted versions of the international trade network, there is a negative dependence between local clustering coefficients and node degrees. This means that the network displays negative assortativity, that is countries with few connections tend to link to highly-connected hubs.

In Section \ref{ModelSection} the model is developed: networks are studied for a finite and an infinite number of reinforcements in the system, and a simulated example is provided. We discuss the empirical results in Section \ref{EmpiricalSection} and extensions of the basic model in Section \ref{Extensions}. Some conclusions and directions of investigation are reported in Section \ref{ConclusionsSection}.

%%%%%%%%%%%%%%%%%%%%%%%%%%%%%%%%%%%%%%%%%%%%%%%%%%%%%%%%%%%%%%%%%%%%%%%%%%%%
%%%%%%%%%%%%%%%%%%%%%%%%%%%%%%%%%%%%%%%%%%%%%%%%%%%%%%%%%%%%%%%%%%%%%%%%%%%%
%%%%%%%%%%%%%%%%%%%%%%%%%%%%%%%%%%%%%%%%%%%%%%%%%%%%%%%%%%%%%%%%%%%%%%%%%%%%
%%%%%%%%%%%%%%%%%%%%%%%%%%%%%%%%%%%%%%%%%%%%%%%%%%%%%%%%%%%%%%%%%%%%%%%%%%%%
\section{Reinforced Urn Model Development}\label{ModelSection}
Consider a directed weighted network $G=(V,E,w^0)$, where $V$ and $E$ are, respectively, the set of vertices and edges, and $w^0:E\rightarrow\mathbb R$ are the initial weights associated to each edge. Take the stochastic process $\{X\}$ to be a Reinforced Urn Process as introduced in \citet{MSW00}, a random walk on Polya urns. We denote $\{X\} \in RUP(V,U,q)$, where $V=\{1,2,\dots,n\}$, the vertex set, is the countable state space of $X_t$. $U=\{U_1,\dots,U_n\}$ is a collection of Polya urns, each associated to a vertex of $G$. The urn associated to vertex $i$ is $U_i=\{w^0_{ij},j=1,\dots,n\}$, where $w^0_{ij}$ is the initial number of balls of colour $j$ in $U_i$. Finally, the law of motion $q:V\times V \rightarrow V$ is responsible for the movement of $\{X\}$ among the vertices: if $X_t=v_1$, a ball is extracted from $U_{v_1}$ and is replaced in $U_{v_1}$ with $s$ additional balls of the same colour; if, say, the extracted ball is $v_2$, then $X_{t+1}=q(v_1,v_2)$. Without loss of generality, we assume $q(i,j)=j$ for all $i,j \in V$. 

After $m$ steps of $\{X\}$, we will have an updated edge weights of $G$. The initial and final (after $m$ steps) network can be equivalently represented by the weighted adjacency matrices $W^0=(w^0_{ij})$ and $W^m=(w^m_{ij})$, where the $ij$-element of the matrix is greater than 0 if the $(i,j)$ edge is in the edge set $E$. Note that the adjacency matrices do not need to be symmetric since the networks are directed, and non-zero entries in the main diagonal 
indicate the presence of loops. We can also have a countably infinite number of nodes, without any substantial change in the results that follow. The starting point of the RUP is the \textit{root} of the network, denoted $v_0$, and, to assure that the RUP comes back to the origin an infinite number of times a.s., we impose, following \citet{MSW00} and \citet{CHM13}, for all $m$, a recurrency condition of the form

$$ \lim_{t\rightarrow+\infty} \prod_{i=0}^t \frac{\sum_{j\neq v_0}w^m_{ij}}{\sum_{j}w^m_{ij}} = 0.$$

Let $m_{ij}$ be the number of times we observe $(X_{t},X_{t+1})=(i,j)$ for $t=1,\dots,m-1$, that is the number of times the directed edge $(i,j)$ is traversed, and let $m_i=\sum_j m_{ij}$ be the number of times vertex $i$ is traversed by the RUP. The probability of observing the network $(w^m_{ij})$ after $m$ steps of the RUP, for $s=1$, is 
\ba
P(W^m = (w^m_{ij})) &=& \prod_{i=1}^n \left\{ \frac{\prod_{j=1}^n \prod_{k=0}^{m_{ij}-1} (w^0_{ij}+k)} {\prod_{k=0}^{m_i-1} (\sum_j w^0_{ij}+k)} \right\}\nn\\
%= \prod_{i=1}^n \left\{ \frac{\prod_{j=1}^n \frac{\Gamma(w^0_{ij}+m_{ij})}{\Gamma(w^0_{ij})}} {\frac{\Gamma(\sum_j w^0_{ij}+m_i)}{\Gamma(\sum_j w^0_{ij})}} \right\} \nn\\
&=& \prod_{i=1}^n \frac{B(\{w^0_{ij}+m_{ij}\}_{j=1}^n)}{B(\{w^0_{ij}\}_{j=1}^n)}\nn\\
&=& \prod_{i=1}^n \frac{B(\{w^m_{ij}\}_{j=1}^n)}{B(\{w^0_{ij}\}_{j=1}^n)}, \label{EqWm}
\ea

\noindent where $B(\{\cdot\}_{j=1}^n)$ is the $n$-variate Beta function. From \citet{MSW00}, $\{X\}$ is Markov exchangeable in the sense of \citet{DF80}, since its law is invariant under permutations of the visited vertices that keep constant the initial states and the transitions. %From \citet{FLPR02}, 
This property of $\{X\}$ is equivalent to row exchangeability %(partial exchangeability in the sense of \citealt{F37} and \citealt{F59}) 
of the matrix of the successor states, that is the matrix whose $(i,j)$-element denotes the value of the stochastic process immediately after the $j$-th visit to  vertex $i$. The adjacency matrix $W^m$ can be recovered deterministically from the matrix of successor states of $\{X\}$, and it retains the row exchangeability property. For this reason, it is of no surprise that Equation \eqref{EqWm} is invariant under row permutation. 
Furthermore, the probability after $m$ RUP steps is invariant under the same permutation of rows and columns, a property known in the literature as \textit{joint exchangeability} (\citealt{K05}) of $W^m$. Networks with joint exchangeable adjacency matrices are also known as \textit{exchangeable graphs} (\citealt{DJ08,OR15}). This intuitively means that the probability law associated to the network is invariant under \textit{relabeling} of the nodes.  Therefore, the probability of seeing a particular network depends only on which patterns occur in the network and how often (how many edges, triangles, five-stars, etc.), but not on where in the network they occur. 

%Therefore, the Aldous-Hoover theorem for weakly exchangeable arrays can be applied to exchangeable graphs:

Some network structures of no interest can be removed by appropriate choices of the initial weight composition. For instance, 
setting $w^0_{ii}=0$ for all $i \in V$ excludes self-loops, and $w^0_{ij}w^0_{jk}w^0_{ki}=0$ for all $i \neq j \neq k$ excludes triangles and focus on trees. The choice $w^0_{i,j}=0$ for all $i$ and for $j \notin \{1,i+1\}$ reduces the model to the RUP on the integer line of \citet{MSW00}. 

The initial set of edges $E$ can be modified, for instance, by removing (i.e. by setting to zero the edge weight) those edges $(i,j)$ 
having normalized weights $w_{ij}^m$ below a fixed threshold $\xi$. This defines a many-to-one transformation from the matrix $W^m$ to a new matrix $W^{m,\xi}$ and
\ba
P(W^{m,\xi}=(w^m_{ij})) &=& \sum_{L \in \mathcal L} \prod_{i=1}^n \left\{ \left[\prod_{\{j:w^m_{ij}\neq0\}} \prod_{k=0}^{m_{ij}-1}(w_{ij}^0+k)\right] \cdot \left[\frac{\prod_{\{j:w^m_{ij}=0\}} \prod_{k=0}^{l_{ij}-1}(w_{ij}^0+k)}{\prod_{k=0}^{\sum_j(m_{ij}+l_{ij})-1} \left(\sum_j w_{ij}^0+k\right)}\right] \right\} \nn\\
&=& \sum_{L \in \mathcal L} \prod_{i=1}^n \frac{B\left(\{w^m_{ij}\}_{j:w^m_{ij}\neq0},\{w_{ij}^0+l_{ij}\}_{j:w^m_{ij}=0}\right)} {B\left(\{w^0_{ij}\}_{j=1}^n\right)}, \label{EqWmxi}
\ea

\noindent where $m_{ij}=\max\{w_{ij}^m-w_{ij}^0,0\}$, $m_i=\sum_j m_{ij}$ and 
\ba
\mathcal L &=& \left\{ L=(l_{ij}) \left|\ \left(\frac{l_{ij}+w_{ij}^0}{m_i+\sum_k( w_{ik}^0 + l_{ik})}<\xi,\ \forall j:\ w_{ij}^m=0\right) \text{ and } \right.\right.\nn\\ 
&& \ \ \ \ \ \ \ \ \ \ \ \ \ \left.\left.\left(\frac{w_{ij}^m}{m_i+\sum_k( w_{ik}^0 + l_{ik})}\geq\xi,\ l_{ij}=0,\ \forall j:\ w_{ij}^m\neq0\right),\ i,j = 1,\dots,n \right.\right\} \nn,
\ea

\noindent under the constraint that $\sum_{i,j}l_{ij} = m - \sum_i m_i$. 
The set $\mathcal L$ identifies all the $W^m$s that correspond to the same observed $W^{m,\xi}$: the generic truncated edge $(i,j)$ is increased, beyond $w_{ij}^0$, by all those quantities $l_{ij}$ that keep the normalized weights of the truncated edges under the threshold and keep the normalized weights of the un-truncated edges above or at the threshold. Finally, $\mathcal L$ is completely characterize by the constraint that the total number of added weights on all edges, relative to the initial configuration, should correspond to the number of RUP steps. 
Note that when $\sum_i{m_i}=m$, there is no edge truncation, $\mathcal L$ contains only a matrix of zeros and Equation \eqref{EqWmxi} simplifies to \eqref{EqWm}.

Restrict now the analysis to the generic vertex $i$ of the network represented in Equation \eqref{EqWm}. From Stirling's approximation, the Gamma function can be written as
$$\Gamma(\alpha z + \beta) \approx \sqrt{2\pi}(\alpha z)^{\alpha z + \beta-1/2}e^{-\alpha z}$$
\noindent for fixed real numbers $\alpha$ and $\beta$ and for the argument $z$ \textit{large enough} in absolute value. From the approximation above, we can derive the approximate right-tail behaviour of conditional and joint distributions of the weights. The joint weights distribution in the right tail (for large values of all the coordinates) can be approximated as
$$
(w_{i1}^m,\dots,w_{in}^m) \propto B(\{w^m_{ij}\}_{j=1}^n)
\propto \frac{\prod_{j=1}^n (w^m_{ij})^{w^m_{ij}-1/2} }{\left(\sum_{j=1}^n w^m_{ij}\right)^{\sum_j w^m_{ij}-1/2}}.
$$

On the other hand, the probability mass function of the weight $w_{ij}^m$, conditional on all the other edges, in the right tail is proportional to 

\ba
w_{ij}^m|\{w_{ik}^m\}_{k\neq j} &\propto& B\left(w^m_{ij},\sum_{k\neq j}w^m_{ik}\right)
\propto (w^m_{ij})^{-\sum_{k\neq j}w^m_{ik}} \nn\\
&\propto& Yule\left(w^m_{ij};\sum_{k\neq j}w^m_{ik} - 1\right), \nn
\ea

\noindent where $Yule$ denotes the Yule-Simon discrete distribution with probability mass function $f(\cdot,\rho) = \rho B(\cdot,\rho+1)$. In words, the weight conditional distribution has a power law right tail with an edge-dependent parameterization. In particular, the decay rate depends negatively 
on the weights of the other edges of the vertex, creating \textit{competition} among the edges of a common node. It is also possible to 
derive the tail behaviour of the joint conditional distributions of subsets of weights: for instance, for large $w_{ij_1}^m$ and $w_{ij_2}^m$:
$$
w_{ij_1}^m,w_{ij_2}^m|\{w_{ik}^m\}_{k\notin \{j_1,j_2\}} \propto B(w_{ij_1}^m,w_{ij_2}^m) (w^m_{ij_1}+w^m_{ij_2})^{-\sum_{k\notin \{j_1,j_2\}}w^m_{ik}}
$$
\noindent and, more generally, for large $w_{ij_1}^m,\dots,w_{ij_h}^m$:
$$
w_{ij_1}^m,\dots,w_{ij_h}^m|\{w_{ik}^m\}_{k\notin \{j_1,\dots,j_h\}} \propto B\left(\{w_{ij_l}^m\}_{l=1}^h\right) \left(\sum_{l=1}^h w^m_{ij_l}\right)^{-\sum_{k\notin \{j_1,\dots,j_h\}}w^m_{ik}}.
$$

Now define $\tilde{w}^m_{ij} \propto w^m_{ij}$, normalized to have $\sum_j\tilde{w}^m_{ij}=1$. When the number of steps $m$ goes to infinity, the superscript $m$ is removed. It is possible to 
derive the joint distribution of $\tilde{W}$, an equivalence class of adjacency matrices $W=(w_{ij})$ with equal $\tilde{w}_{ij}$. It is well known from \citet{A69} that the colours' proportions of a Polya urn converge in probability to a Dirichlet random variable, therefore
\ba
P(\tilde W = (\tilde w_{ij})) = \prod_{i=1}^n \left\{ \frac{\prod_{j=1}^{n-1} (\tilde w_{ij})^{w^0_{ij}/s-1} (1-\sum_{j=1}^{n-1}\tilde w_{ij})^{w^0_{in}/s-1}}{B(\{w^0_{ij}/s\}_{j=1}^n)} \right\}. \label{EqWtilde}
\ea
Note that the equivalence classes so determined are prediction consistent, since $$P(X_{m+1}=j|X_1,\dots,X_m=i) = \tilde w_{ij},$$ \noindent with all networks in the same class sharing the same prediction on $\{X\}$.

Furthermore, we can remove those edges $(i,j)$ for which $\tilde w_{ij}$ is below a threshold $\xi$, where 
\ba
P((i,j) \text{ removed in } \tilde W) = I_\xi\left( \frac{w^0_{ij}}{s},\frac{\sum_{k\neq j}w^0_{ik}}{s} \right) = \frac{B\left(\xi; \frac{w^0_{ij}}{s},\frac{\sum_{k\neq j}w^0_{ik}}{s} \right)}{B\left( \frac{w^0_{ij}}{s},\frac{\sum_{k\neq j}w^0_{ik}}{s} \right)},\label{EqRemove}
\ea
\noindent and $I_\xi(\cdot,\cdot)$ and $B(\xi;\cdot,\cdot)$ are, respectively, the regularized and the incomplete Beta functions. It can be shown that the difference between the densities of two weighted directed networks $\tilde W_1$ and $\tilde W_2$ corresponding to the same network after the truncation of edges below $\xi$ is bounded above by
$$P(\tilde W_1)-P(\tilde W_2) \leq  \prod_{i=1}^n \left\{ \frac{\prod_{\{j:\tilde w_{ij}\neq0\}}\tilde w_{ij}^{w_{ij}^0/s-1}}{B\left(\left\{w^0_{ij}/s\right\}_{j=1}^n\right)} \right\} \cdot \xi^{\sum_{i=1}^n \sum_{\{j:\tilde w_{ij}=0\}}(w^0_{ij}/s-1)},$$

\noindent converging to 0 as $\xi$ vanishes. Asymptotically in $m$, the transition 
from weighted directed networks to unweighted directed networks, and therefore from $\tilde W$ to the unweighted directed adjacency matrix $A=(a_{ij})$, is such that marginally 
$$a_{ij} \sim Bern\left(1-I_\xi\left( \frac{w^0_{ij}}{s},\frac{\sum_{k\neq j}w^0_{ik}}{s} \right)\right),$$
\noindent and the likelihood of a specific configuration $A=(a_{ij})$ can be written as
\ba
P(A = (a_{ij})) = \prod_{i=1}^n \int_{\Omega_i} 1_{C_i} \frac{\prod_{j=1}^{n-1}\tilde w_{ij}^{w^0_{ij}/s-1} (1-\sum_{j\neq n}\tilde w_{ij})^{w^0_{in}/s-1}}{B(\{w^0_{ij}/s\}_{j=1}^n)} \prod_{j=1}^{n-1} d\tilde w_{ij},\label{EqA}
\ea
\noindent where $\Omega_i = [\xi a_{ij}, \xi + (1-\xi)a_{ij}]^{\bigotimes_{j=1}^{n-1}}$ and $C_i$ is the condition $(-1)^{a_{in}}\sum_{j=1}^{n-1}\tilde w_{ij} \geq (-1)^{a_{in}} (1-\xi)$, for $i=1,\dots,n$. Finally, the unweighted undirected adjacency matrix $B$ is obtained from $A$, considering that $B=(b_{ij})$ is such that, on the main diagonal, $b_{ii}=a_{ii}$, and 
off-diagonally, 
$$b_{ij}=\left\{\begin{array}{cc} 1 & \text{ if } a_{ij}=1 \text{ or } a_{ji}=1 \\ 
                                  0 & \text{ otherwise} \end{array}\right..$$ 
Therefore, marginally, for $i \neq j$
$$b_{ij} = b_{ji} \sim Bern\left(1-I_\xi\left( \frac{w^0_{ij}}{s},\frac{\sum_{k\neq j}w^0_{ik}}{s} \right) I_\xi\left( \frac{w^0_{ji}}{s},\frac{\sum_{k\neq i}w^0_{jk}}{s} \right)\right),$$
\noindent and the probability of a random configuration of $B$ is
\ba
P(B=(b_{ij})) = \sum_{\{i \neq j: b_{ij}=1\}} \sum_{(b_{ij},b_{ji}) \in \{(0,1),(1,0),(1,1)\}} P(A=(b_{ij})).\label{EqB}
\ea

\begin{figure}[ht]
\begin{center}
\caption{Network configurations, with likelihood computations detailed in the Exercise of Section \ref{ModelSection}.}
\label{Fig1}
\subfloat[$W^0$]{
\begin{tikzpicture}[->,>=stealth',shorten >=1pt,auto,node distance=3cm,
  thick,main node/.style={circle,fill=blue!20,draw,font=\sffamily\Large\bfseries}]

  \node[main node] (v1) {$v_1$};
  \node[main node] (v2) [below of=v1] {$v_2$};
  \node[main node] (v3) [below right of=v1] {$v_3$};

  \path[every node/.style={font=\sffamily\small}]
    (v1) edge node [left] {2} (v2)
         edge [bend left] node [left] {1} (v3)
    (v2) edge node [left] {2} (v3)
         edge [bend left] node [left] {1} (v1)
    (v3) edge node [left] {2} (v1)
         edge [bend left] node [left] {1} (v2);
\end{tikzpicture}\label{Fig1a}}
\subfloat[$W^3_1$]{
\begin{tikzpicture}[->,>=stealth',shorten >=1pt,auto,node distance=3cm,
  thick,main node/.style={circle,fill=blue!20,draw,font=\sffamily\Large\bfseries}]

  \node[main node] (v1) {$v_1$};
  \node[main node] (v2) [below of=v1] {$v_2$};
  \node[main node] (v3) [below right of=v1] {$v_3$};

  \path[every node/.style={font=\sffamily\small}]
    (v1) edge node [left] {3} (v2)
         edge [bend left] node [left] {1} (v3)
    (v2) edge node [left] {3} (v3)
         edge [bend left] node [left] {1} (v1)
    (v3) edge node [left] {3} (v1)
         edge [bend left] node [left] {1} (v2);
\end{tikzpicture}\label{Fig1b}}
\subfloat[$W^3_2$]{
\begin{tikzpicture}[->,>=stealth',shorten >=1pt,auto,node distance=3cm,
  thick,main node/.style={circle,fill=blue!20,draw,font=\sffamily\Large\bfseries}]

  \node[main node] (v1) {$v_1$};
  \node[main node] (v2) [below of=v1] {$v_2$};
  \node[main node] (v3) [below right of=v1] {$v_3$};

  \path[every node/.style={font=\sffamily\small}]
    (v1) edge node [left] {2} (v2)
         edge [bend left] node [left] {2} (v3)
    (v2) edge node [left] {2} (v3)
         edge [bend left] node [left] {2} (v1)
    (v3) edge node [left] {2} (v1)
         edge [bend left] node [left] {2} (v2);
\end{tikzpicture}\label{Fig1c}}\\
\subfloat[$A$]{
\begin{tikzpicture}[->,>=stealth',shorten >=1pt,auto,node distance=3cm,
  thick,main node/.style={circle,fill=blue!20,draw,font=\sffamily\Large\bfseries}]

  \node[main node] (v1) {$v_1$};
  \node[main node] (v2) [below of=v1] {$v_2$};
  \node[main node] (v3) [below right of=v1] {$v_3$};

  \path[every node/.style={font=\sffamily\small}]
    (v1) edge node [left] {}(v2)
    (v2) edge node [left] {}(v3)
         edge [bend left] node [left] {}(v1)
    (v3) edge [bend left] node [left] {}(v2);
\end{tikzpicture}\label{Fig1d}}
\subfloat[$B$]{
\begin{tikzpicture}[-,>=stealth',shorten >=1pt,auto,node distance=3cm,
  thick,main node/.style={circle,fill=blue!20,draw,font=\sffamily\Large\bfseries}]

  \node[main node] (v1) {$v_1$};
  \node[main node] (v2) [below of=v1] {$v_2$};
  \node[main node] (v3) [below right of=v1] {$v_3$};

  \path[every node/.style={font=\sffamily\small}]
    (v1) edge node [] {}(v2)
    (v2) edge node [] {}(v3);
\end{tikzpicture}\label{Fig1e}}
\subfloat[$W^{9,\xi}$]{
\begin{tikzpicture}[->,>=stealth',shorten >=1pt,auto,node distance=3cm,
  thick,main node/.style={circle,fill=blue!20,draw,font=\sffamily\Large\bfseries}]

  \node[main node] (v1) {$v_1$};
  \node[main node] (v2) [below of=v1] {$v_2$};
  \node[main node] (v3) [below right of=v1] {$v_3$};

  \path[every node/.style={font=\sffamily\small}]
    (v1) edge node [left] {4} (v2)
    (v2) edge node [left] {4} (v3)
    (v3) edge node [left] {4} (v1);
\end{tikzpicture}\label{Fig1f}}
\end{center}
\end{figure}
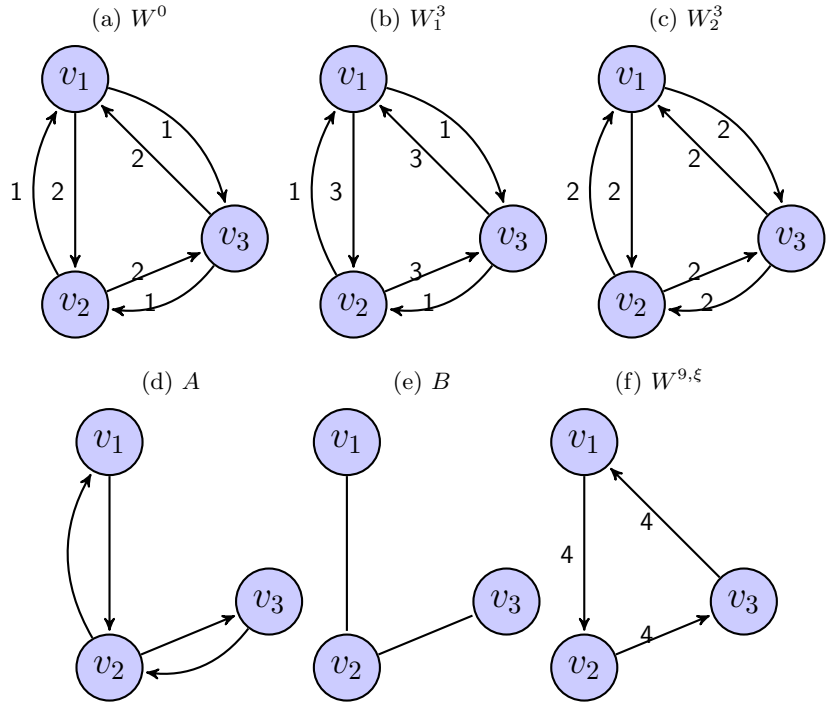

\noindent \textbf{Example:} 
Consider the weighted directed network $W^0$ in Figure \ref{Fig1a}. Suppose $v_1$ is the root vertex, where the RUP starts its walk. After $m=3$ RUP steps, the probability of observing the configuration in Figure \ref{Fig1b} is, from Equation \eqref{EqWm}, $$P(W^3=W^3_1) = \left(\frac{B(3,1)}{B(2,1)}\right)^3 = 0.2963,$$ \noindent whilst the network in Figure \ref{Fig1c} has a much lower probability 
$$P(W^3=W^3_2) = \left(\frac{B(2,2)}{B(2,1)}\right)^3 = 0.0370,$$ \noindent 
since it is less likely to traverse the triangle clockwise. 
From Equation \eqref{EqWtilde}, when $m\rightarrow\infty$ and the reinforcement is $s=1$, the two equivalence classes $\tilde W_1$ and $\tilde W_2$, to which $W^3_1$ and $W^3_2$ respectively belong, have a probability density function of 3.375 and 1, 
respectively. Furthermore, from Equation \eqref{EqRemove}, fixing the threshold $\xi=0.2$, we can compute $P((v_1,v_3) \text{ removed in } \tilde W)=I_{0.2}(1,2)=0.36$ and $P((v_3,v_1) \text{ removed in } \tilde W)=I_{0.2}(2,1)=0.04$. Starting from $W^0$, we will observe the unweighted directed network $A$ in Figure \ref{Fig1d} with probability given in \eqref{EqA}
\ba
P(A = (a_{ij})) &=& P(\tilde w_{12}\geq\xi,\tilde w_{13}<\xi)P(\tilde w_{21}\geq\xi,\tilde w_{23}\geq\xi)P(\tilde w_{32}\geq\xi,\tilde w_{31}<\xi)\nn\\
&=& I_{\xi}(\tilde w_{13}^0,\tilde w_{12}^0)\left(I_{1-\xi}(\tilde w_{21}^0,\tilde w_{23}^0)-I_{\xi}(\tilde w_{21}^0,\tilde w_{23}^0) \right) I_{\xi}(\tilde w_{31}^0,\tilde w_{32}^0) \nn\\
&=& 0.00864.\nn
\ea
\noindent The unweighted undirected $B$ in Figure \ref{Fig1e}, according to \eqref{EqB}, is observed with probability
\ba
P(B = (b_{ij})) &=& P(\tilde w_{12}\geq\xi,\tilde w_{13}<\xi) \left\{P(\tilde w_{21}\geq\xi,\tilde w_{23}\geq\xi)+\right.\nn\\
&& \left.P(\tilde w_{21}\geq\xi,\tilde w_{23}<\xi)+P(\tilde w_{21}<\xi,\tilde w_{23}\geq\xi)\right\} P(\tilde w_{32}\geq\xi,\tilde w_{31}<\xi)\nn\\
&=& I_{\xi}(\tilde w_{13}^0,\tilde w_{12}^0)I_{\xi}(\tilde w_{31}^0,\tilde w_{32}^0) \nn\\
&=& 0.0144. \nn
\ea
\noindent Starting from $W^0$ in Figure \ref{Fig1a}, configuration in Figure \ref{Fig1f} is observed after 9 RUP steps with probability depending on $\xi$. The truncated edges are $(1,3)$, $(3,2)$ and $(2,1)$ and, for $\xi=6/13$, 
$$
\mathcal L = \left\{ \begin{pmatrix} 0 & 0 & 1 \\ 1 & 0 & 0 \\ 0 & 1 & 0 \end{pmatrix}, \begin{pmatrix} 0 & 0 & 2 \\ 1 & 0 & 0 \\ 0 & 0 & 0 \end{pmatrix}, \begin{pmatrix} 0 & 0 & 2 \\ 0 & 0 & 0 \\ 0 & 1 & 0 \end{pmatrix}, \begin{pmatrix} 0 & 0 & 1 \\ 2 & 0 & 0 \\ 0 & 0 & 0 \end{pmatrix}, \begin{pmatrix} 0 & 0 & 0 \\ 2 & 0 & 0 \\ 0 & 1 & 0 \end{pmatrix}, \begin{pmatrix} 0 & 0 & 1 \\ 0 & 0 & 0 \\ 0 & 2 & 0 \end{pmatrix}, \begin{pmatrix} 0 & 0 & 0 \\ 1 & 0 & 0 \\ 0 & 2 & 0 \end{pmatrix} \right\},
$$
\noindent with the probability of the configuration in \ref{Fig1f} given in Equation \eqref{EqWmxi} and equal to 
$$
P(W^{9,6/13}=(w_{ij}^9)) = \frac{B(4,2)}{B(2,1)^3}\left\{ B(4,2)^2 + 6 B(4,1) B(4,3) \right\} = 0.011.
$$

%%%%%%%%%%%%%%%%%%%%%%%%%%%%%%%%%%%%%%%%%%%%%%%%%%%%%%%%%%%%%%%%%%%%%%%%%%%%
%%%%%%%%%%%%%%%%%%%%%%%%%%%%%%%%%%%%%%%%%%%%%%%%%%%%%%%%%%%%%%%%%%%%%%%%%%%%
%%%%%%%%%%%%%%%%%%%%%%%%%%%%%%%%%%%%%%%%%%%%%%%%%%%%%%%%%%%%%%%%%%%%%%%%%%%%
%%%%%%%%%%%%%%%%%%%%%%%%%%%%%%%%%%%%%%%%%%%%%%%%%%%%%%%%%%%%%%%%%%%%%%%%%%%%
\section{International Trade Network Analysis}\label{EmpiricalSection}
Each year of the DOT data corresponds to a realized network. Each row indicates the exports of a country towards other countries in the world. For comparability among years, each matrix always includes the same countries for all the years, so that
countries with null imports and exports are included in the analysis. Note that all the diagonal elements of the matrix are null, indicating absence of self loops in the network. This is implied by the nature of the problem which focuses on international trade, and is not a limitation of the methodology. Still, it would be possible to extend the analysis to network with self loops, for instance including in the study internal consumption data. We want to be able to update our beliefs about a trading relationship between two countries in the future, based on the observations of trading relationships that has happened in the past. Moreover, we are willing to reinforce the probability of observing a specific export composition. 

The directional trade flows in the most recent available year constitutes the $186 \times 186$ matrix $\tilde W$. Different choices for the prior network can be made, and we opt for $W^0$ implied by the bilateral trade in the remotest year, that is 1948. All zero elements in $W^0$ have been set equal to 
$$\xi = \min_{i,j:\ w^0_{ij} \neq 0}w^0_{ij},$$ 
\noindent to allow the creation of a new trading relationships among countries that had no trading relationship in 1948. 

For several choices of the reinforcement parameter $s \in \{1,2,\dots,200\}$, we compute the probability of observing the present export composition, starting from the trade flows in 1948. Following a maximum likelihood approach, we then estimate $s$ with the value of $s$ that maximizes this probability. Since our likelihood is available in closed form, we can also add a prior on $s$ and conduct Bayesian posterior inference. The maximum likelihood estimated reinforcement is $\hat s = 5$, and it coincides with the Maximum a Posteriori estimator in case of an assumed uniform prior on the reinforcement $s$. 
Finally we run a robustness analysis on the estimate of $s$ by conducting a prior sensitivity study. This robustness  check involves substituting the original uniform prior distribution on $s$ with different exponential prior distributions having rate parameters 0.01, 0.05, 0.5, 1 and 2 respectively. The differences in the posterior densities are negligible, with the posterior mode being  always equal to 5 for any prior. From the other posterior summaries of $s$ in Table \ref{TableS} we note that the likelihood is the dominating part in the parameter posterior inference, since the posterior mean only slightly decreases as the mean of the prior distribution decreases (i.e. as we go down the table). Also the 95\% credible intervals show a light switch to lower values. Posterior skewness remains negative for all prior scenarios, with little differences mainly due to a slight increase in the posterior variance. Finally, the excess kurtosis is always negative, but relatively close to the one obtained in the Gaussian case. 

\begin{figure}[ht]
\begin{center}
\caption{(a) Estimated international trade network country-specific reinforcements $s$. (b) Distribution of the differences between the actual and forecasted number of partnerships, where the forecast is the average over 10000 simulated networks infinitely reinforced by the RUP. (c) Kolmogorov-Smirnov (KS) distances between actual and forecasted exports, for each country. (d) KS distances between the two forecasted scenarios for each country, where the two forecasts average 10000 predictions, starting from the initial networks of, respectively, 1988 and 1991.}
\subfloat[]{\includegraphics[width=5cm]{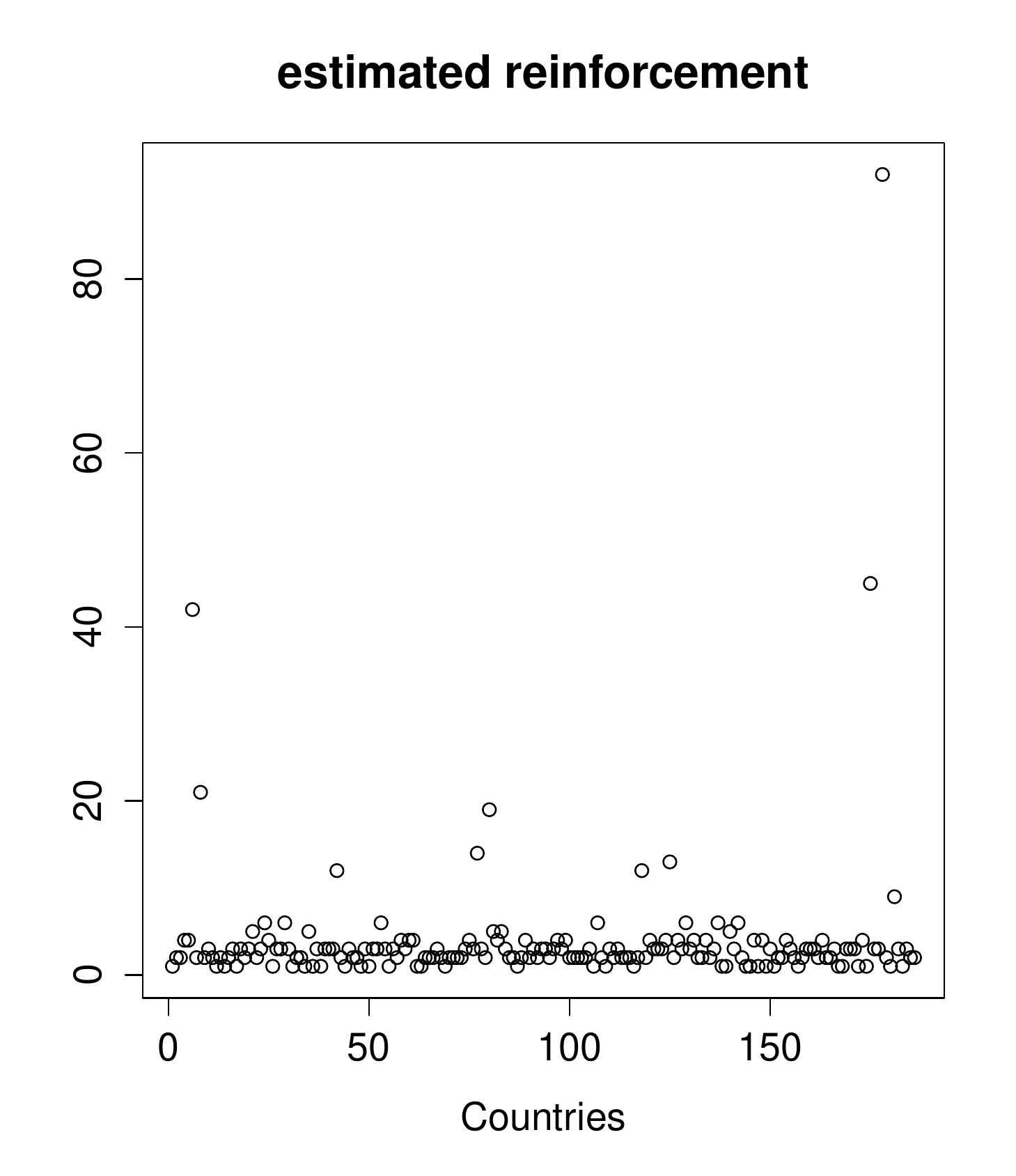}\label{Fig2a}}
\subfloat[]{\includegraphics[width=5cm]{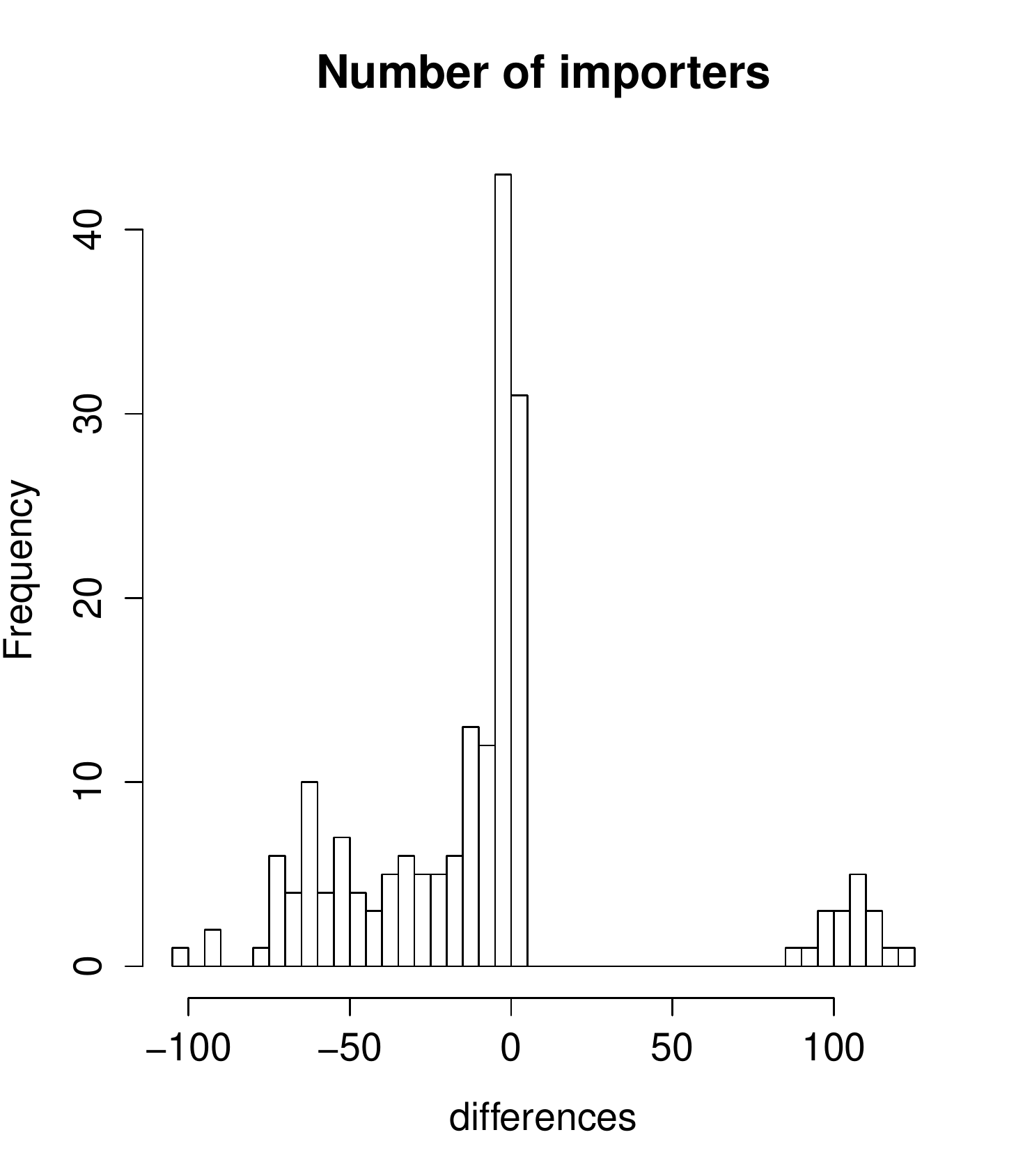}\label{Fig2b}}\\
\subfloat[]{\includegraphics[width=5cm]{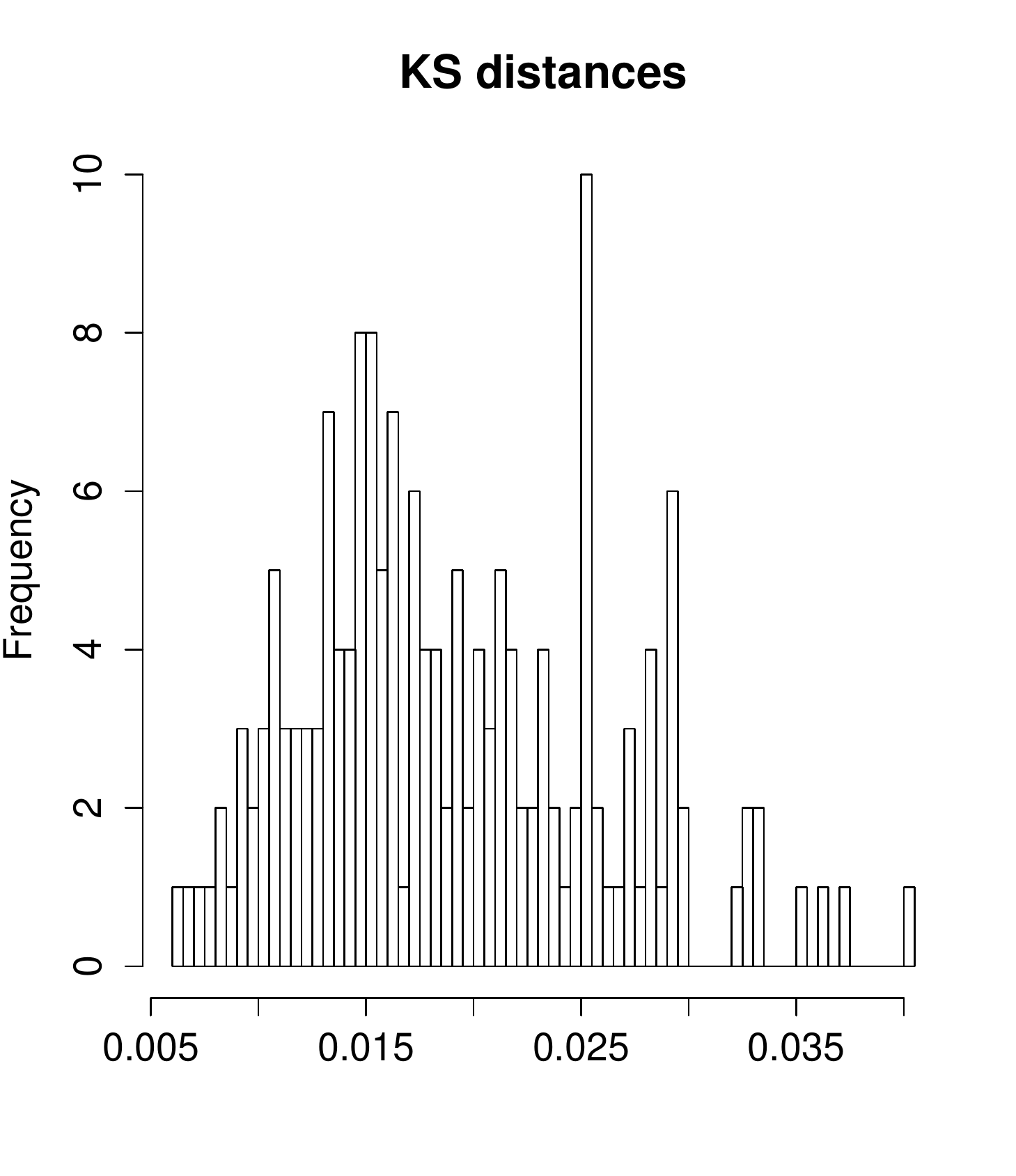}\label{Fig2c}}
\subfloat[]{\includegraphics[width=5cm]{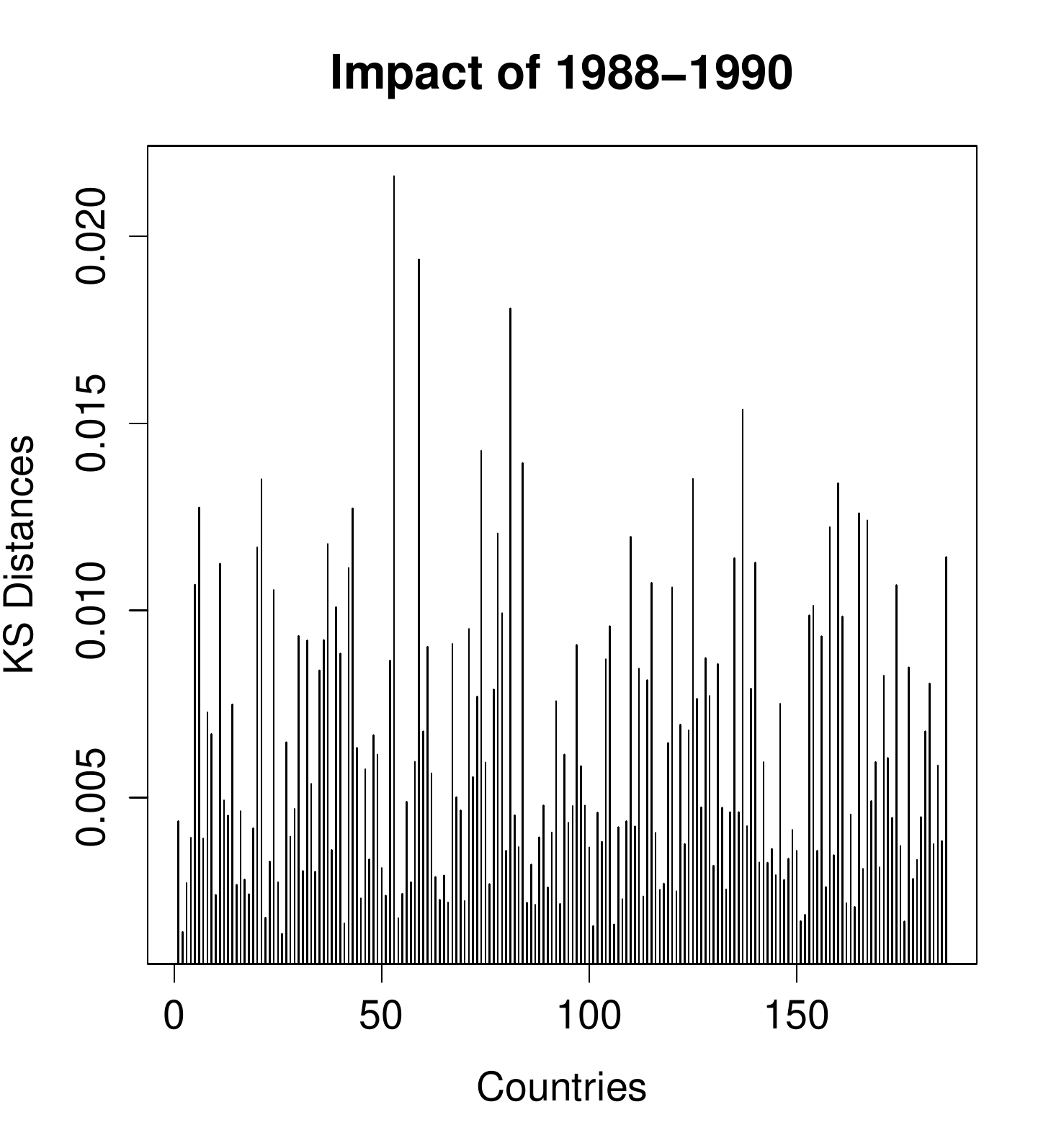}\label{Fig2d}}
\label{Fig2}
\end{center}
\end{figure}

A different reinforcement parameter can be estimated for each vertex of the network, allowing for  country-specific reinforcements. In this way, we account for countries having different abilities to consolidate their partnerships. The estimated reinforcements $s$ are shown in Figure \ref{Fig2a}. A great majority of the countries, 169 out of 186, has a reinforcement between 1 and 5 included, and for almost all of them $s$ is less than or equal to 21. The notable exceptions are USA, United Kingdom and Argentina, with an estimated $s$ of, respectively, 92, 45, and 42. 

At this point, we forecast the network, starting from year 2000, after an infinite number of RUP steps, given reinforcements estimated as above. Again, all the null elements of $\tilde W$ are fixed to a threshold equal to the minimum non-null element of $\tilde W$, thus allowing for new trading partnerships. $M=10000$ forecasts are simulated, and $\tilde W^{sim}$, the average of all simulated matrices, is taken as expected forecast. All the elements in $\tilde W^{sim}$ below the threshold are then fixed to 0. The distribution of the differences between the actual and forecasted number of partnerships is plotted in the histogram of Figure \ref{Fig2b}. The peak around 0 highlights all those 24 countries that do not change their set of trading partners. There are 51 countries that widen their set of partners, but the high left skewness points out a forecasted concentration of partners for 111 countries. In particular, for Canada, United Kingdom and USA it is forecasted a concentration of trade flows towards some of their major importers, for instance an increase in the exports from Canada and USA to Japan and from United Kingdom to Ireland. Note also a cluster of countries, mostly from Africa and Central America, which considerably increase their exports. Finally, for each country, we compute the Kolmogorov-Smirnov (KS) distance between actual and forecasted exports, and we plot the distances in Figure \ref{Fig2c}. It is possible to scan the results at the desired level of detail: as an example, the country presenting the maximum KS distance is Lithuania, for which the exports towards France and Norway are forecasted to, respectively, decrease from 5.27\% to 4.14\% and increase from 1.65\% to 2.75\%.

\begin{figure}[ht]
\begin{center}
\caption{Average forecast, out of 10000 predicted networks, of the network representing the import-export relations among the Group of Eight countries. The edge widths are reported proportional to the corresponding weights.}
\includegraphics[width=10cm]{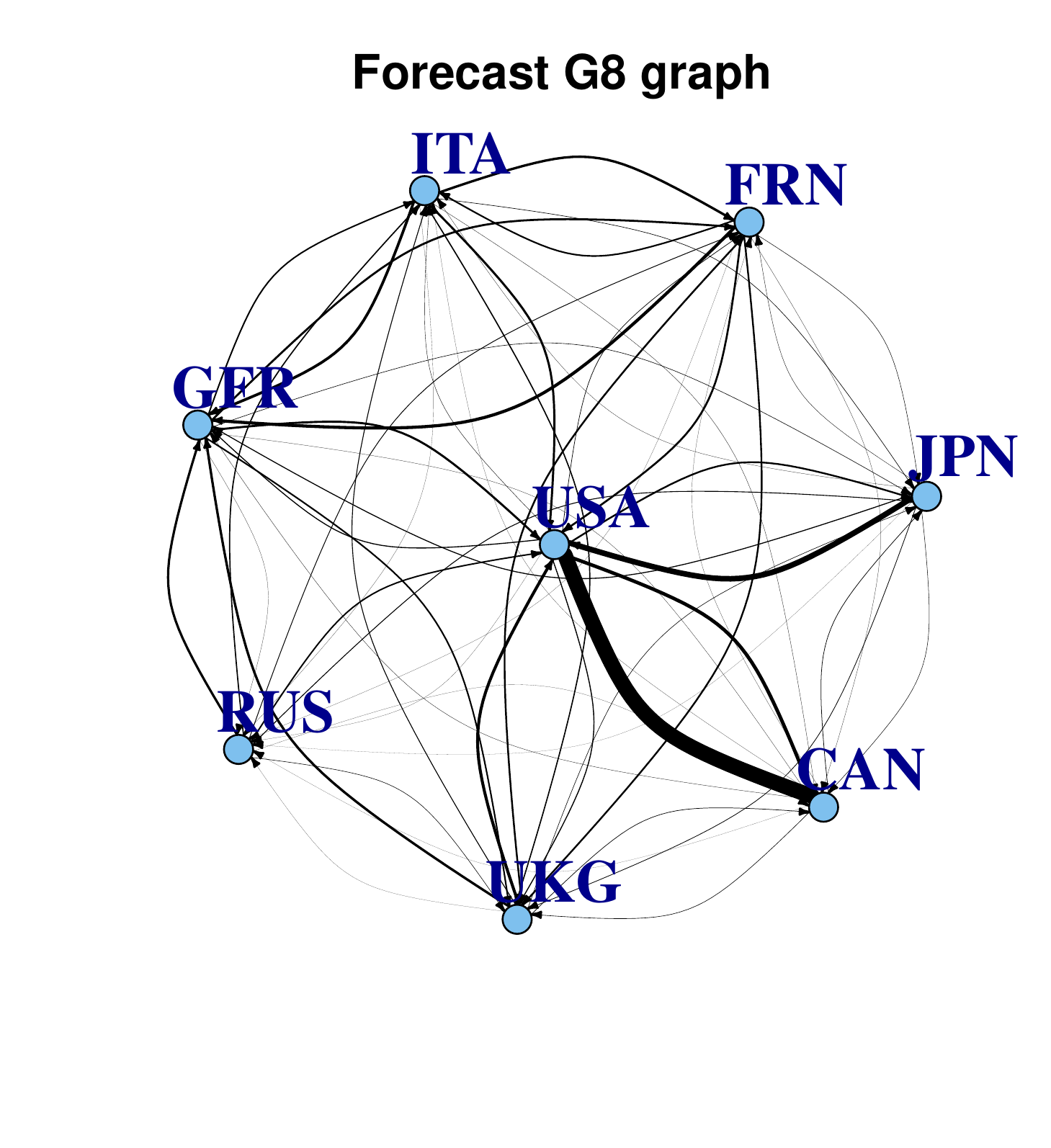}
\label{Fig3}
\end{center}
\end{figure}

Selecting two different starting points, $W_0$s, observed at two different years, we can repeat the procedure above and, informally, understand how the events in between the two starting points
can affect the prediction. As an example, we compute $M$ predictions starting from the initial network configurations observed in 1988 and 1991. In Figure \ref{Fig2d} we report, in terms of KS distances between the two forecasted scenarios for each country, the impact of the events between 1988 and 1990. Finally, in Figure \ref{Fig3}, we plot the average forecast of the network representing the import-export relations among the Group of Eight countries (G8). The edge widths are reported proportional to weights, to better identify stronger international relationship, as between United States and Canada and between United States and Japan.

%%%%%%%%%%%%%%%%%%%%%%%%%%%%%%%%%%%%%%%%%%%%%%%%%%%%%%%%%%%%%%%%%%%%%%%%%%%%
%%%%%%%%%%%%%%%%%%%%%%%%%%%%%%%%%%%%%%%%%%%%%%%%%%%%%%%%%%%%%%%%%%%%%%%%%%%%
%%%%%%%%%%%%%%%%%%%%%%%%%%%%%%%%%%%%%%%%%%%%%%%%%%%%%%%%%%%%%%%%%%%%%%%%%%%%
%%%%%%%%%%%%%%%%%%%%%%%%%%%%%%%%%%%%%%%%%%%%%%%%%%%%%%%%%%%%%%%%%%%%%%%%%%%%
\section{Extensions}\label{Extensions}
In Section \ref{ModelSection} we developed our model based on the system of Polya urns, and derived implications in terms of power-law behaviour of the strength distribution. Different reinforcement parameters $s$ for different countries allow incorporating, in the model, countries with various propensities to trade and consolidate their partnerships. A more general parameterization of the reinforcement $s$ permits the treatment of other empirical regularities observed in the international trade network.

In particular, we can define $s_{ij}$ the reinforcement associated to balls in urn $i$ that link country (vertex) $i$ with country $j$, extending the previous framework from country-specific reinforcement to bilateral flow-specific reinforcement. 
Specifically, let
$$s_{ij} = \alpha_i + \beta_{ij} \sum_k (\tilde w_{jk}+ \tilde w_{kj}) + \gamma_{ij} \sum_k (\tilde w_{ik}+\tilde w_{ki})(\tilde w_{jk}+\tilde w_{kj}).$$
In the above formula, a negative value of $\beta_{ij}$ captures the negative assortativity, that is the empirical phenomenon for which countries with few connections tend to link to highly-connected hubs, since vertices $j$ with higher trade flows (sum of in-strength and out-strength) are disadvantaged. 
Similarly, a positive $\gamma_{ij}$ accounts for the \textit{path-shortening} tendency to observe closed triangles, favoring partnerships with countries having common partners. Finally, different levels of the threshold $\xi$ defined in Section \ref{ModelSection}, for which edges with weight lower than $\xi$ are finally removed, control for the global sparsity of the network. A higher $\xi$ induces lower sparsity, since there will be a higher proportion of edges with strictly positive weights. Fixing $\beta_{ij}=\gamma_{ij}=\xi=0$ reduces the model to the one discussed in Section \ref{ModelSection}. 

We emphasize that likelihood evaluation and simulation is feasible also in this extended setting, and therefore exact inference can be performed. Still, to avoid the computational inefficiencies implicit in numerical likelihood evaluation with multidimensional parameters, we can follow \citet{MO15} and estimate $\alpha_i$, $\beta_{ij}$, $\gamma_{ij}$ and $\xi$ through an Approximate Bayesian Computation (ABC) approach (\citealt{Diggle1984,Rubin1984,Tavare1997}). When ABC is applied to network models, the original dataset corresponds to the observed network $G$. 
The parameter values $\alpha_i$, $\beta_{ij}$, $\gamma_{ij}$ and $\xi$ are sampled from the prior distributions, typically taken to be non-informative. With the generated configuration, a sample of networks $G_1, \ldots, G_T$ is obtained, exploiting the fact that it is easy to generate a graph configuration from the RUP path once a set of parameter values is specified. The observed network $G$ is compared to the model-generated networks $G_1, \ldots, G_T$ using appropriate summary statistics. For a univariate summary statistic, a natural choice is to use the degree distribution (or the strength distribution in case of weighted networks): we extract the degree distributions $p_i(k)$ for the model-generated networks $G_i$ and compare each of them to the degree distribution $p(k)$ of the observed network $G$. 
One possible distance for performing this comparison is the two-sample Kolmogorov-Smirnov (KS) test statistics, which compares the empirical cumulative distribution functions of $p_i(k)$ and  $p(k)$ to determine how close they are to one another. The proposal $(\alpha_i,\beta_{ij},\gamma_{ij},\xi)$ is accepted whenever the value of the KS test statistic is less than a pre-specified critical value, or, equivalently, whenever the $p$-value associated with the test is greater than the corresponding critical value $p^*$. The collection of accepted proposals form the approximation of the parameters' posterior distribution. Note that if the value of $p^*$ is very low, we end up accepting many of the proposed values and, in the limit of $p^* \to 0$, our estimated posterior distribution simply ends up recovering the prior distribution. The  pseudo-code of the algorithm is shown in Algorithm \ref{fig:code}.

If we are interested in more than one summary statistic, the random forest approach of \citet{Breiman2001} can be used, following again the methodology in \citet{MO15}. In particular, adopting the  random forest  approach, we could jointly consider summary statistics as node degree, node strength, edge strength, local and global clustering (weighted and unweighted), path lengths, eigenvector centrality, closeness, modularity, number of communities. Indeed, the principle behind the ABC random forest approach is to rely on a classifier that can handle a relatively large number of summary statistics, instead of selecting among the statistics or relying on the expertise of the user to pick up a relatively small number of statistics relevant for the problem at hand. 

Finally, \cite{Pudlo2014} propose an ABC model selection procedure with random forests choosing among highly complex models covered by ABC algorithms. Adopting a similar methodology and following again \citet{MO15}, model selection among different urn-based graphs can be performed with the ABC random forest approach.

%%%%%%%%%%%%%%%%%%%%%%%%%%%%%%%%%%%%%%%%%%%%%%%%%%%%%%%%%%%%%%%%%%%%%%%%%%%%
%%%%%%%%%%%%%%%%%%%%%%%%%%%%%%%%%%%%%%%%%%%%%%%%%%%%%%%%%%%%%%%%%%%%%%%%%%%%
%%%%%%%%%%%%%%%%%%%%%%%%%%%%%%%%%%%%%%%%%%%%%%%%%%%%%%%%%%%%%%%%%%%%%%%%%%%%
%%%%%%%%%%%%%%%%%%%%%%%%%%%%%%%%%%%%%%%%%%%%%%%%%%%%%%%%%%%%%%%%%%%%%%%%%%%%
\section{Conclusions and further directions} \label{ConclusionsSection}
To theoretically justify in a unified framework the empirical regularities of the international trade network, we propose a model for networks based on a system of Polya urns, in which the walk of a Reinforcement Urn Process feeds the system, generating weighted edges (bilateral trade flows among countries). A basic model for directed weighted networks is provided, and probability laws governing different kinds of networks (weighted, unweighted, directed and undirected) are derived through different notions of equivalence classes of networks. Model properties are studied, and applications to simulated data and to the motivating problem are provided. 
Our  urn-based network model  can account for preferential attachment, negative assortativity, path-shortening and global sparsity
and, contrary to \citet{BA99}, we provide a closed form expression for the likelihood that allows likelihood and posterior based inference.
The main limitation of our approach is that model dynamics is not treated. To study the model behaviour as the number of countries involved (vertices) increases, it could be worth investigating the special form that the limiting density of \citet{LS06} assumes in the proposed model, through the derivation of the distribution of finite adjacency submatrices of networks with countably infinite vertices. 
To investigate model dynamics over time, a specific transition law of discrete time graph-valued Markov chains could be 
introduced, exploiting the general setting of  \citet{C15}.
%%%%%%%%%%%%%%%%%%%%%%%%%%%%%%%%%%%%%%%%%%%%%%%%%%%%%%%%%%%%%%%%%%%%%%%%%%%%
%%%%%%%%%%%%%%%%%%%%%%%%%%%%%%%%%%%%%%%%%%%%%%%%%%%%%%%%%%%%%%%%%%%%%%%%%%%%
%%%%%%%%%%%%%%%%%%%%%%%%%%%%%%%%%%%%%%%%%%%%%%%%%%%%%%%%%%%%%%%%%%%%%%%%%%%%
%%%%%%%%%%%%%%%%%%%%%%%%%%%%%%%%%%%%%%%%%%%%%%%%%%%%%%%%%%%%%%%%%%%%%%%%%%%%
\section*{Acknowledgements}
We thank prof. Jukka-Pekka Onnela for fruitful discussions on network modelling. 
The Swiss based authors gratefully acknowledge Swiss National Science Foundation financial support.

\singlespacing
\bibliographystyle{apalike}
%\nocite{MSI12}
\bibliography{refGraph}

\begin{thebibliography}{}

\bibitem[Athreya, 1969]{A69}
Athreya, K. (1969).
\newblock {On a Characteristic Property of P\'olya's Urn}.
\newblock {\em Stud. Sci. Math. Hung.}, 4:31--35.

\bibitem[Barabasi and Albert, 1999]{BA99}
Barabasi, A. and Albert, R. (1999).
\newblock {Emergence of Scaling in Random Networks}.
\newblock {\em {Science}}, 286:509--512.

\bibitem[Barrat et~al., 2004]{B04}
Barrat, A., Bartel\'emy, M., Pastor-Satorras, R., and Vespignani, A. (2004).
\newblock {The architecture of complex weighted networks}.
\newblock {\em Proceedings of the National Academy of Sciences},
  101:3747--3752.

\bibitem[Bhattacharya et~al., 2007]{BMM07}
Bhattacharya, K., Mukherjee, G., and Manna, S. (2007).
\newblock {The international trade network}.
\newblock In {\em {Econophysics of Markets and Business Networks}}, pages
  139--147.

\bibitem[Breiger, 1981]{B81}
Breiger, R. (1981).
\newblock {Structures of economic interdependence among nations}.
\newblock In {\em {Continuities in structural inquiry}}, pages 353--380.

\bibitem[Breiman, 2001]{Breiman2001}
Breiman, L. (2001).
\newblock Random forests.
\newblock {\em Machine Learning}, 45(1):5--32.

\bibitem[Cirillo et~al., 2013]{CHM13}
Cirillo, P., Husler, J., and Muliere, P. (2013).
\newblock {Alarm Systems and Catastrophes from a Diverse Point of View}.
\newblock {\em {Methodology and Computing in Applied Probability}},
  15:821--839.

\bibitem[Crane, 2015]{C15}
Crane, H. (2015).
\newblock {Dynamic random networks and their graph limits}.
\newblock {\em {Annals of Applied Probability}}.

\bibitem[Diaconis and Freedman, 1980]{DF80}
Diaconis, P. and Freedman, D. (1980).
\newblock {De Finetti's Theorem for Markov Chains}.
\newblock {\em {Annals of Probability}}, 8:115--130.

\bibitem[Diaconis and Janson, 2008]{DJ08}
Diaconis, P. and Janson, S. (2008).
\newblock {Graph limits and exchangeable random graphs}.
\newblock {\em {Rendiconti di Matematica, Serie VII}}, 28:33--61.

\bibitem[Diggle, 1984]{Diggle1984}
Diggle, P. (1984).
\newblock {Monte Carlo Methods of Inference for Implicit Statistical Models}.
\newblock {\em {J. of the Royal Statistical Society, Series B}}, 46:193--227.

\bibitem[Fagiolo et~al., 2008]{FRS08}
Fagiolo, G., Reyes, J., and Schiavo, S. (2008).
\newblock {On the topological properties of the world trade web: A weighted
  network analysis}.
\newblock {\em Physica A: Statistical Mechanics and its Applications},
  387:3868--3873.

\bibitem[Fagiolo et~al., 2009]{FRS09}
Fagiolo, G., Reyes, J., and Schiavo, S. (2009).
\newblock {World-trade web: Topological properties, dynamics, and evolution}.
\newblock {\em Physical Review E}, 79:36115.

\bibitem[Fisher, 2003]{F03}
Fisher, S. (2003).
\newblock {Globalization and its challenges}.
\newblock {\em American Economic Review}, 93:1--30.

\bibitem[Garlaschelli and Loffredo, 2004]{GL04}
Garlaschelli, D. and Loffredo, M. (2004).
\newblock {Fitness-dependent topological properties of the world trade web}.
\newblock {\em Physical review letters}, 93:188701.

\bibitem[Garlaschelli and Loffredo, 2005]{GL05}
Garlaschelli, D. and Loffredo, M. (2005).
\newblock {Structure and evolution of the world trade network}.
\newblock {\em Physica A: Statistical Mechanics and its Applications},
  355:138--144.

\bibitem[Kallenberg, 2005]{K05}
Kallenberg, O. (2005).
\newblock {\em Probabilistic Symmetries and Invariance Principles}.
\newblock Springer, New York.

\bibitem[Lovasz and Szegedy, 2006]{LS06}
Lovasz, L. and Szegedy, B. (2006).
\newblock {Limits of Dense Graph Sequences}.
\newblock {\em {Journal of Combinatorial Theory, Series B}}, 96:933--957.

\bibitem[Mira and Onnela, 2015]{MO15}
Mira, A. and Onnela, J.-P. (2015).
\newblock {Statistical inference on large-scale mechanistic network models}.

\bibitem[Muliere et~al., 2000]{MSW00}
Muliere, P., Secchi, P., and Walker, S. (2000).
\newblock {Urn schemes and reinforced random walks}.
\newblock {\em Stochastic Processes and their Applications}, 28:59--78.

\bibitem[Newman, 2005]{N06}
Newman, M. (2005).
\newblock {Power laws, Pareto distributions and Zipf's law}.
\newblock {\em {Contemporary physics}}, 46:323--351.

\bibitem[Opsahl and Panzarasa, 2009]{OP09}
Opsahl, T. and Panzarasa, P. (2009).
\newblock {Clustering in weighted networks}.
\newblock {\em Social Networks}, 31:155--163.

\bibitem[Orbanz and Roy, 2015]{OR15}
Orbanz, P. and Roy, D. (2015).
\newblock {Bayesian models of graphs, arrays and other exchangeable random
  structures}.
\newblock {\em {IEEE Transactions on Pattern Analysis and Machine
  Intelligence}}, 37:437--461.

\bibitem[Price, 1976]{P76}
Price, D. (1976).
\newblock {A general theory of bibliometric and other cumulative advantage
  processes}.
\newblock {\em {Journal of the American Society for Information Science}},
  27:292--306.

\bibitem[Pudlo et~al., 2014]{Pudlo2014}
Pudlo, P., Marin, J., Estoup, A., Cornuet, J., Gautier, M., and Robert, C.
  (2014).
\newblock {ABC model choice via random forests}.

\bibitem[Rauch, 2001]{R01}
Rauch, J. (2001).
\newblock {Business and social networks in international trade}.
\newblock {\em Journal of Economic Literature}, 39:1177--1203.

\bibitem[Robins et~al., 2007]{R07}
Robins, G., Pattison, P., Kalish, Y., and Lusher, D. (2007).
\newblock {An introduction to exponential random graphs (p*) models for social
  networks}.
\newblock {\em {Social Networks}}, 29:173--191.

\bibitem[Rubin, 1984]{Rubin1984}
Rubin, D. (1984).
\newblock {Bayesianly Justifiable and Relevant Frequency Calculations for the
  Applies Statistician}.
\newblock {\em {The Annals of Statistics}}, 12:1151--1172.

\bibitem[Serrano and Bogun\'a, 2003]{SB03}
Serrano, M. and Bogun\'a, M. (2003).
\newblock {Topology of the world trade web}.
\newblock {\em Physical Review E}, 68:15101.

\bibitem[Simon, 1955]{S55}
Simon, H. (1955).
\newblock {On a Class of Skew Distribution Functions}.
\newblock {\em {Biometrika}}, 42:425--440.

\bibitem[Snyder and Kick, 1979]{SK79}
Snyder, D. and Kick, E. (1979).
\newblock {Structural position in the world system and economic growth,
  1955-1970: A multiple-network analysis of transnational interactions}.
\newblock {\em American Journal of Sociology}, 84:1096--1126.

\bibitem[Tavare et~al., 1997]{Tavare1997}
Tavare, S., Balding, D., Griffiths, R., and Donnelly, P. (1997).
\newblock {Inferring Coalescence Times From DNA Sequence Data}.
\newblock {\em {Genetics}}, 145:505--518.

\bibitem[Watts and Strogatz, 1998]{WS98}
Watts, D. and Strogatz, S. (1998).
\newblock {Collective dynamics of small-world networks}.
\newblock {\em Nature}, 393:440--442.

\end{thebibliography}
\normalsize

\newpage
\section*{Figures}

\begin{table}
\caption{Posterior summaries of the reinforcement parameter $s$, for different prior distributions}
\label{TableS}
\begin{center}
\begin{tabular}{lccccc} \hline
Prior				& Mean		& 95\% interval			& Variance		& Skewness		& Kurtosis\\\hline
$U([0,200])$  & 4.8485 	& [2.1295,5.8976] 	& 1.0031			& -0.8468			& 2.6729  \\
$Exp(0.01)$	& 4.8484 	& [2.1294,5.8976] 	& 1.0032			& -0.8466			& 2.6725  \\
$Exp(0.05)$	& 4.8480 	& [2.1293,5.8976] 	& 1.0035			& -0.8461			& 2.6708  \\
$Exp(0.5)$	& 4.8435 	& [2.1283,5.8970] 	& 1.0074			& -0.8397			& 2.6516  \\
$Exp(1)$	  & 4.8384 	& [2.1271,5.8964] 	& 1.0116			& -0.8327			& 2.6305  \\
$Exp(2)$	  & 4.8283 	& [2.1248,5.8952] 	& 1.0200			& -0.8184			& 2.5886  \\\hline
\end{tabular}
\end{center}
\end{table}

\begin{algorithm}
\begin{algorithmic}
\For {iteration $t = 1, 2, \ldots, T$}
	\State Draw proposal $\alpha_i$, $\beta_{ij}$, $\gamma_{ij}$ and $\xi$ from the prior distribution
	\State Generate graph $G_i$ from the RUP model, given extracted parameters
	\State Set $d_i = \rho(S(G_i), S(G_0))$, where $S$ is some summary statistic and $\rho$ is a distance
\EndFor
\State Set $\epsilon = P_1(d)$ (first percentile)
\For {iteration $t = 1, 2, \ldots, T$}
	\If{ $d_i \le \epsilon$ }  Accept $\alpha_i$, $\beta_{ij}$, $\gamma_{ij}$ and $\xi$
	\EndIf
\EndFor
\end{algorithmic}
\caption{ABC rejection sampler for RUP network models.}
\label{fig:code}
\end{algorithm}

\end{document}